\documentclass[11pt]{article}
\usepackage[centertags]{amsmath}
\usepackage{amsfonts}
\usepackage{amssymb}
\usepackage{amsthm}
\usepackage{newlfont}
\usepackage{graphicx}

\newfont{\bb}{msbm10 at 10pt}

\begin{document}
\begin{center}
\large{\textbf{Plane symmetric cosmological models}}
\\
\vspace{4mm}
Anil Kumar Yadav$^{1}$, Ahmad T Ali$^{2}$, Saibal Ray$^{3}$,\\ Farook Rahaman$^{4}$ and Arkapriya Mallick$^{4}$\\
\vspace{4mm}
$^{1}$Department of Physics\\ United College of Engineering and Research, Greater Noida 201306, India\\
{\it {E-mail: abanilyadav@yahoo.co.in}}\\
\vspace{4mm}
$^{2}$Department of Mathematics\\ Faculty of Science, King Abdul Aziz University, PO Box 80203
Jeddah, 21589, Saudi Arabia\\
{\it {E-mail: ahmadtawfik95@gmail.com}}\\
\vspace{4mm}
$^{3}$Department of Physics\\ Government College of Engineering and Ceramic Technology, Kolkata 700010, West Bengal, India\\  
{\it {E-mail: saibal@associates.iucaa.in}}\\
\vspace{4mm}
$^{4}$Department of Mathematics\\ Jadavpur University, Kolkata 700032, West Bengal,
India\\ 
{\it {E-mail: rahaman@associates.iucaa.in}}\\
\end{center}
\vspace{4mm}
\begin{abstract}
In the present work, we execute the Lie symmetry analysis on the Einstein-Maxwell field equations in the plane symmetric spacetime. Under the background of the plane symmetric space-time we compute the Lie point symmetries, perform the similarity reductions and obtain exact solutions in connection to the evolutionary scenario of the universe. The special feature of the study is that it deals with the electromagnetic energy of the inhomogeneous universe through the non-vanishing component of electromagnetic field tensor $F_{12}$ and assumes that the free gravitational field is of Petrov type-II non-degenerate. We have found that the electromagnetic field tensor is positive and increasing function of time. To validate the solution set, we examine with detailed discussions several physical as well as geometrical features of a specific sub-case of the model.
\end{abstract}



\section{Introduction}
The Friedman-Robertson-Walker (FRW)~\cite{Walker1937} cosmological model describes a universe that configures the smoothness of the cosmic spacetime. On the other hand, the current astrophysical observations also have exhibited that the distribution of matter is isotropic and the geometry of the present universe is in the form of spherical symmetry. However, it is also quite clear that at the early stage of evolution the universe could not have such a smoothed picture. Keeping this aspect in mind, in the present study we confine ourselves to construct the model of an  inhomogeneous universe by considering the metric polynomials as function of the space and time both.

At theoretical point of view, the inhomogeneous cosmological models are too impartant mainly due to the following two reasons: (i) less validity of isotropic distribution of matter closure to the big bang impulse, and (ii) Perturbation present in the standard cosmological model. Firstly, Tolman~\cite{Tolman1934} and thereafter other scientists, viz. Bondi~\cite{Bondi1947}, Taub~\cite{Taub1951,Taub1956}, Tomimura~\cite{Tomimura1978}, Szekeres~\cite{Szekeres1975} and Senovilla~\cite{Senovilla1990} have time to time constructed variety of plane symmetric inhomogeneous cosmological models through the exact solution of Einstein's general relativistic field equations. Later on, Ruis and Senovilla~\cite{Senovilla1990} have presented a significant work which is singularity-free inhomogeneous cosmological model. Further Bali and Tyagi~\cite{Bali1990} and Pradhan et al.~\cite{Pradhan2007a,Pradhan2010} have extended their studies on the plane symmetric inhomogeneous cosmological model under the Einstein-Maxwell space-time. Recently, Pradhan et al.~\cite{Pradhan2009}, Yadav~\cite{Yadav2010}, Ali and Yadav~\cite{Ali2014a} and Ali et al.~\cite{Ali2014b} have also investigated various inhomogeneous cosmological models taking into consideration of different physical aspects.

In 1983, Zeldovich et al.~\cite{Zeldovich1983} had been investigated the effect of magnetic field on galactic scale that gives clue for the importance of magnetic field for different astrophysical phenomenon. Harrison~\cite{Harrison1973} has also discussed some applications of magnetic field in cosmology which motivate us to include magnetic field in the right side of Einstein's equation. In the recent past, the results of WMAP data are in favour of anisotropic modelling of universe. Some authors~\cite{Asseo1987,Hughston1970} have speculated that the occurance of a primordial magnetic field is bounded to some anisotropic models of universe spacially in case of  B-I,  II, III, $VI_{0}$ and $VII_{0}$. In this connection Zeldovich~\cite{Zeldovich1970} and Barrow~\cite{Barrow1997} have investigated that the magnetized anisotropic pressure dominates over the evolution caused by shear anisotropy. However, the existence of such field may possible at the end of an inflationary epoch as argue by several authors~\cite{Turner1988,Quashnock1989,Dolgov1993}. All the above studies indicate that the magnetized anisotropic models of universe play significant role in the evolving process of galaxies as well as stellar systems.

The Symmetry analysis method is a powerful tool for solving these equations~\cite{Bluman1989,Ovsyannikov1982,Olver1993,Zhang2011}. These methods have been successfully applied in the area of theoretical physics, in particular quantum mechanics, fluid dynamics and particle physics~\cite{Baumann2000,Ibragimov1985}. We have elaborated this technique in the area of relativistic cosmology. Some applications of the Lie point symmetry analysis methods are discussed by Ali et al.~\cite{Ali2014a,Ali2014b}. For more details of the Lie groups, one may consult the following works~\cite{Baumann2000,Olver1995,Stephani1989} which deal with several applications including reduction of order of PDEs, development of similarity solution and generating new solutions from known ones. The classification of group-invariant solutions of differential equations by means of the so-called optimal system is one of the main applications of Lie group analysis of differential equations. The method was first conceived by Ovsiannikov~\cite{Ovsyannikov1982}. In 1985, Ibragimov~\cite{Ibragimov1985} has given some examples of optimal system in his book. Later on, Olver~\cite{Olver1993} has summarized some interesting discussion on similarity solution based on optimal systems.

In the present study, therefore, our goal is to search for an effective method for solving the system of nonlinear PDEs. This methodology we employ for accelerating universe in the plane symmetric space-time filled with non-exotic matter and electromagnetic fluid under general relativistic background. The outline of our study is as follows: in Sec. 2 we introduce the mathematical modeling of the accelerating universe in the plane symmetric space-time whereas Sec. 3 deals with the solution of the field equations. In Sec. 4 we provide some studies of physical and geometrical properties of our models and its validity. A few comments with a short discussion are presented in Sec. 5.\\

\section{The Einstein-Maxwell spacetime geometry}
The line element considered here is as follows
\begin{equation} \label{u2-10}
ds^{2}=A^{2} \left(dx^{2}-dt^{2}\right)+B^{2} dy^{2}+C^{2} dz^{2},
\end{equation}
where the metric potentials $A$, $B$ and $C$ are functions of the spatial and temporal coordinates $x$ and $t$ both.

The usual form of energy-momentum tensor is given by
\begin{equation}
\label{eq2} T^{\iota}_{\nu} = (\rho + p)v_{\nu}v^{\iota} + p g^{\iota}_{\nu} + E^{\iota}_{\nu},
\end{equation}
where $E^{\iota}_{\nu}$ is the electromagnetic field and it is read as
\begin{equation}
\label{eq3} E^{\iota}_{\nu} = \bar{\mu}\left[\zeta_{\ell}\zeta^{\ell}(v_{\nu}v^{\iota} +
\frac{1}{2}g^{\iota}_{\nu}) - \zeta_{\nu}\zeta^{\iota}\right].
\end{equation}

In above equations the parameters $\rho$, $p$, $v^{\iota}$, $\bar{\mu}$ and $\zeta_{\ell}$ are the energy density, isotropic pressure, flow vector, magnetic permeability and magnetic flux vector respectively where 
\begin{equation}
\label{eq5} \zeta_{\ell} = \frac{1}{\bar{\mu}}~~ ^*F_{\iota \nu}v^{j}.
\end{equation}

Again, the dual electro-magnetic field tensor$^{}${Synge1960} is given by
\begin{equation}
\label{eq6} ^*F_{\nu\iota} = \frac{\sqrt-g}{2}\epsilon_{\nu \iota k\ell} F^{k\ell}.
\end{equation}

Here, we assume the current is flowing along $z$ axis and $F_{12}$ is the only non-vanishing component of electro-magnetic field tensor and the Maxwell equations are read as  
\begin{equation}
\label{eq7} \Biggl[\frac{1}{\bar{\mu}} F^{\iota \nu}\Biggr]_{;\iota} = 0,
\end{equation}
where the semicolon (;) stands for their usual meaning, i.e. the matter creation through non-zero left hand side is possible while conserving the over all energy and momentum.

The Maxwell equation (\ref{eq7}) leads to
$\frac{\partial }{\partial x}\left[\frac{F_{12}(x,t) C(x,t)}{\bar{\mu}(x,t)B(x,t)}\right]\\=0$
which requires that
\begin{equation}
\label{eq8} \frac{F_{12}(x,t)C(x,t)}{\bar{\mu}(x,t)B(x,t)}=f(t),
\end{equation}
where $f(t)$ is an arbitrary function of $t$ and we assume the component $F_{12}$ of the electromagnetic field and the magnetic permeability $\bar{\mu}$ as a functions of $x$ and $t$ both.

The Einstein field equations are read as
\begin{equation}
\label{eq9} R^{j}_{i} - \frac{1}{2} R g^{j}_{i} = - 8\pi T^{j}_{i},
\end{equation}

Solving Eq. (1) with Eq. (\ref{eq9}), one obtain
\begin{equation}  \label{u210}
E_1=\frac{B_{xt}}{B}+\frac{C_{xt}}{C}-\frac{A_t}{A}\left(\frac{B_x}{B}+\frac{C_x}{C}\right)
    -\frac{A_x}{A}\left(\frac{B_t}{B}+\frac{C_t}{C}\right)=0,
\end{equation}

\begin{eqnarray}  \label{u211}
E_2=\frac{B_t C_t-B_x C_x}{BC}+\frac{B_{tt}}{B}+\frac{C_{xx}}{C}+\frac{A_{xx}-A_{tt}}{A}
+\frac{A_t}{A}\left(\frac{A_t}{A}-\frac{B_t}{B}-\frac{C_t}{C}\right)\nonumber\\-\frac{A_x}{A}\left(\frac{A_x}{A}+\frac{B_x}{B}+\frac{C_x}{C}\right)=0,
\end{eqnarray}

\begin{eqnarray}  \label{u212}
\chi A^2 p(x,t)=\frac{B_{xx}-2 B_{tt}}{2 B}-\frac{C_{tt}}{2 C}+\frac{A_{xx}-A_{tt}}{2 A}+ \frac{B_x C_x-B_t C_t}{2 B C}
+\frac{A_t}{2A}\left(\frac{A_t}{A}+\frac{B_t}{B}+\frac{C_t}{C}\right)- \nonumber\\ \frac{A_x}{2A}\left(\frac{A_x}{A}-\frac{B_x}{B}-\frac{C_x}{C}\right),
\end{eqnarray}

\begin{eqnarray}  \label{u213}
\chi A^2 \rho(x,t)=\frac{C_{tt}-2 C_{xx}}{2 C}-\frac{B_{xx}}{2 B}+\frac{A_{xx}-A_{tt}}{2 A}+\frac{3\big(B_t C_t-B_x\,C_x\big)}{2 B C}
+\nonumber\\ \frac{A_t}{2 A}\left(\frac{A_t}{A}+\frac{B_t}{B}+\frac{C_t}{C}\right)+ \frac{A_x}{2\,A}\left(\frac{A_x}{A}+\frac{B_x}{B}+\frac{C_x}{C}\right),
\end{eqnarray}

\begin{eqnarray}  \label{u214}
\frac{\chi F^2_{12}(x)}{B^2 \bar{\mu}(x,t)}=\frac{B_x C_x-B_t C_t}{B C}-\frac{C_{tt}}{C}-\frac{B_{xx}}{B}+\frac{A_{tt}-A_{xx}}{A}
+\frac{A_t}{A}\left(\frac{C_t}{C}+\frac{B_t}{B}-\frac{A_t}{A}\right)\nonumber\\+\frac{A_x}{A}\left(\frac{A_x}{A}+\frac{B_x}{B}+\frac{C_x}{C}\right).
\end{eqnarray}

The scalar expansion $\Theta$ and shear scalar $\sigma^2$ can be provided as$^{}${dec1,rayc1}:
\begin{equation}  \label{u215-c}
\Theta=\frac{1}{A} \Big(\frac{A_t}{A}+\frac{B_t}{B}+\frac{C_t}{C}\Big),
\end{equation}

\begin{equation}  \label{u215-d}
\begin{array}{ll}
\sigma^2\, =\frac{\Theta^2}{3}-\frac{1}{A^2}\Big(\frac{A_t B_t}{A B}+\frac{A_t C_t}{A C}+
\frac{B_t C_t}{B C}\Big).
\end{array}
\end{equation}

In Eqs. (\ref{u210})-(\ref{u214}), there are five highly non-linear differential equations with six unknown variables, viz. $A$, $B$, $C$, $p$, $\rho$ and $F_{12}^2/\bar{\mu}$. In general, it is impossible to solve these equations without assuming physically reasonable conditions amongst the parameters. In the present situation for the model (\ref{u2-10}), let us assume that  $\Theta \propto \sigma^{1}_{1}$ which leads to the relation between the metric potentials as follows:
\begin{equation}\label{u223}
  \begin{array}{ll}
\frac{2A_t}{A}-\frac{B_t}{B}-\frac{C_t}{C}=3\delta \Big(\frac{A_t}{A}+\frac{B_t}{B}+\frac{C_t}{C}\Big),
  \end{array}
\end{equation}
where $\delta$ is a proportional constant.

Equation (\ref{u223}) can be written in the following convenient form
\begin{equation}\label{u224}
  \begin{array}{ll}
\frac{A_t}{A}=n\Big(\frac{B_t}{B}+\frac{C_t}{C}\Big),
  \end{array}
\end{equation}
where $n=\frac{1+3\delta}{2-3\delta}$.

Integrating it with respect to $t$, we obtain
\begin{equation}\label{u225}
  \begin{array}{ll}
A(x,t)=f(x)B^n(x,t)C^n(x,t),
  \end{array}
\end{equation}
where $f(x)$ is an integration constant and has an arbitrary functional relationship with $x$.

Equations (\ref{u225}), (\ref{u210}) and (\ref{u211}) lead the following equations
\begin{eqnarray}  \label{u210-1}
E_1=\frac{B_{xt}}{B}+\frac{C_{xt}}{C}-2n\left(\frac{B_x B_t}{B^2}+\frac{B_x C_t+B_t C_x}{BC}
    +  \frac{C_x C_t}{C^2}\right)- \frac{f'}{f}\left(\frac{B_t}{B}+\frac{C_t}{C}\right)=0,
\end{eqnarray}

\begin{eqnarray}  \label{u210-2}
E_2=n\left(\frac{B_{xx}}{B}-\frac{C_{tt}}{C}\right)+(1-n)\frac{B_{tt}}{B}+(1+n)\frac{C_{xx}}{C}
                + 2n\left(\frac{B_x^2}{B^2}-\frac{C_x^2}{C^2}\right)
        +(1-2n)\frac{B_t C_t}{B C}-\nonumber\\ (1+2n)\frac{B_x C_x}{B C}- \frac{f'}{f}\left(\frac{B_x}{B}+\frac{C_x}{C}\right)+\frac{f f''-f^{\prime 2}}{f^2}=0.
\end{eqnarray}

Here and in what follows, a prime indicates derivative with respect to the coordinate $x$.\\
\section{The solutions of the Einstein-Maxwell field equations}
If we solve the system of second order nonlinear partial differential equations (NLPDEs) involved in Eqs. (\ref{u210-1})-(\ref{u210-2}),
we shall get the exact solution of the  problem under consideration. Several authors~\cite{Pradhan2007b,Yadav2009,Ali2013} have adopted a simple approach to obtain exact solution by taking into account $B(x,t)=B_1(x) B_2(t)$ and $C(x,t)=C_1(x) C_2(t)$. However, in the present paper we have investigated for a new solution by using the symmetry analysis method~\cite{Ovsyannikov1982,Olver1993,Baumann2000,Bluman2002} and optimal system~\cite{Ovsyannikov1982,Olver1993}. It is to note that a complete description of the methods to solve NLPDEs in the framework of general relativity are available in the following works~\cite{Yadav2014,Ali2014b}.

Thus for metric (1), the components of symmetries are given by
\begin{equation}\label{u37}
\xi_{1}=c_1 x+c_2,~\xi_{2}=c_1 t+c_3,~\eta_{1}=c_4 B,~\eta_{2}=c_5 C,
\end{equation}
where the function $f(x)$ must be taken the following forms
\begin{equation}\label{u37-1}\left\{
                               \begin{array}{ll}
                                 f(x) =c_6 \exp\big[c_7 x\big],\,\,\,\,\,\,\,\,\,\,\,\,\,\,\mathrm{if}\,\,\,\,\,c_1=0, \\
\\
                                 f(x)\,=\,c_8\big(c_1\,x+c_2\big)^{c_9},\,\,\,\,\,\,\,\,\,\mathrm{if}\,\,\,\,\,c_1 \neq 0,
                               \end{array}
                             \right.
\end{equation}
with $c_i$ an arbitrary constant where $i=1,2,...,9$.

The equation (\ref{u37-1}) leads the following optimal systems
\begin{equation}\label{u32-5}\left\{
\begin{array}{ll}
X^{(1)}=X_1+c_4 X_4+c_5 X_5,\\
X^{(2)}=c_2 X_2+X_3+c_4 X_4+c_5 X_5,\\
X^{(3)}=X_2+c_4 X_4+c_5 X_5,\\
X^{(4)}=X_4+c_5 X_5,\\
X^{(5)}=X_5\
\end{array}
\right.
\end{equation}

In the present work our procedure of solving the NLPDEs is similar to the works~\cite{Yadav2014,Ali2014b} but here the source of
the energy-momentum tensor and explicit expressions of all the cosmological parameters are entirely different. Basically in
this study we have presented a model of accelerating universe filled with perfect fluid and electromagnetic field in
the framework of inhomogeneous plane symmetric space-time. To our knowledge, this is the first study that deals
with the inhomogeneous modelling of accelerating universe by taking into account the perfect fluid and electromagnetic
field as the source of matter-energy density.

The characteristic equations for symmetries (\ref{u37}) are read as
\begin{equation}\label{u41}
\frac{dx}{c_1 x+c_2}=\frac{dt}{c_1 t+c_3}=\frac{dB}{c_4 B}=\frac{dC}{c_5 C}.
\end{equation}

From Eq. (\ref{u32-5}), it is to be noted that $c_{1} = c_{3} = 0$ for symmetries $X^{(3)}$, $X^{(4)}$ or $X^{(5)}$. Also Eq. (\ref{u41}) leads to the similarity variable as $\xi=t$, where $B$ and $C$ have functional relations with $t$. Therefore, invariant solutions are possible only for the following two cases:\\

{\bf Case I - Symmetries $X^{(2)}$:} From Eq. (\ref{u41}), with $c_1=0$ and $c_3=1$, we have 
\begin{equation}\label{u42-1}
\begin{array}{ll}
\xi=x+b t,~B(x,t)=\varphi(\xi) \exp[c x],\\C(x,t)=\phi(\xi) \exp[d x],
\end{array}
\end{equation}
where $b=-c_2$, $c=\frac{c_4}{c_2}$ and $d=\frac{c_5}{c_2}$ are arbitrary constants.\\

{\bf Case II - Symmetries $X^{(1)}$:} From Eq. (\ref{u41}), with $c_1=1$ and $c_2=c_3=0$, we have
\begin{equation}\label{u52-1}
\begin{array}{ll}
\xi=\frac{t}{x},~B(x,t)=x^c \varphi(\xi),~C(x,t)=x^d \phi(\xi),
\end{array}
\end{equation}
where $c=c_4$ and $d=c_5$ are arbitrary constants.

However, one can perform mathematical and physical analysis by considerding several subcases under the above two cases and conclude that {\it Case~I} and some of its subcases lead us to the physically interesting and viable solutions. Therefore, to save time as well as space, we shall consider only {\it Case~I} and following subcases in our calculations.

Hence, substitution of the transformations (\ref{u42-1}) in Eqs. (\ref{u210-1})-(\ref{u210-2}) provide 
\begin{eqnarray}\label{u42-2}
\Big[c_7-c+2n(d-c)\Big]\frac{\varphi'}{\varphi}+\Big[c_7-d+2n(c-d)\Big]\frac{\phi'}{\phi}
+2n\Big(\frac{\varphi'}{\varphi}+\frac{\phi'}{\phi}\Big)^2-\frac{\varphi''}{\varphi}-\frac{\phi''}{\phi}=0,
\end{eqnarray}

\begin{eqnarray}\label{u42-3}
\Big[(n-1)b^2-n\Big]\frac{\varphi''}{\varphi}-\Big[n+1-nb^2\Big]\frac{\phi''}{\phi}
+ 2n\Big(\frac{\varphi'^2}{\varphi^2}+\frac{\phi'^2}{\phi^2}\Big)
+\left[2n+1+(2n-1)b^2\right]\frac{\varphi' \phi'}{\varphi \phi}
+\nonumber\\ \Big[c_7+d+2n(c+d)\Big]\frac{\varphi'}{\varphi}
+\Big[c_7+d+2n(c+d)-2a_4\Big]\frac{\phi'}{\phi}+\nonumber\\ d(c-d)+c_7(c+d)+n(c+d)^2=0.
\end{eqnarray}

Eqs. (\ref{u42-2}) and (\ref{u42-3}) being difficult to solve one may consider a special case with $b=-1$. Therefore, subtracting in 
Eq. (\ref{u42-2}) and (\ref{u42-3}), we get
\begin{equation}\label{0u42-4}
\begin{array}{ll}
\frac{\phi'}{\phi}+\frac{(c+d)\varphi'}{(c-d)\varphi}=\alpha_0,
\end{array}
\end{equation}
where $\alpha_0=\frac{d(c_7-d)+c(c_7+d)+n(c+d)^2}{d-c}$.

After integration the above equation with respect to $\xi$, we obtain
\begin{equation}\label{0u42-6}
\begin{array}{ll}
\phi(\xi)=r_1 \varphi^{\alpha_1}(\xi)\exp[\alpha_0 \xi],
\end{array}
\end{equation}
where $\alpha_1=\frac{d+c}{d-c}$ while $r_1$ is an arbitrary integration constant.

The equation (\ref{u42-2}), after using the transformation
\begin{equation}\label{0u42-8}
                               \begin{array}{ll}
\varphi(\xi)=r_2 \exp\Big[\alpha_2 \int \Omega(\xi) d\xi\Big]
                               \end{array}
\end{equation}
and (\ref{0u42-6}), becomes
\begin{equation}\label{0u42-9}
\begin{array}{ll}
\Omega'=\eta_1 \Omega^2+\eta_2 \Omega+\eta_3,
\end{array}
\end{equation}
where
\begin{eqnarray}
\left\{
 \begin{array}{ll}\label{0u42-9-1}
\eta_1=\frac{\alpha_2 \Big[(c^2+3d^2)(c^2+2cd-d^2)+4d^2 \left[(c-d)\alpha_0+(c+d)c_7\right]\Big]}{d(c-d)(c+d)^2},\\

\eta_2=\frac{c^4-8c^2 d^2-4cd^3+3d^4-2\Big[c^3+5c^2d+7cd^2-5d^3\Big]\alpha_0
+8d \alpha_0 \Big[(d-c)\alpha_0-(c+d)c_7\Big]}{2d(c+d)^2}-c_7,\\

\eta_3=\frac{\alpha_0 (c-d) \Big[2c(d+\alpha_0)(d+c_7+\alpha_0)-d(d-c_7+\alpha_0)(d+2\alpha_0)+c^2(3d+c_7+3\alpha_0)\Big]}{2d\alpha_2 (c+d)^2},
\end{array}
\right.
\end{eqnarray}
and $r_2$ is constant while $\Omega(\xi)$ is a new function of $\xi$.

To get solution of the above ordinary differential equation we consider the following special cases: $\eta_1 \neq 0$, $\eta_2=0$ and 
$\eta_3 \neq 0$ so that the general solution to Eq. (\ref{0u42-9}) becomes
\begin{equation}\label{0u42-10-2}
\begin{array}{ll}
\Omega(\theta)=\sqrt{\frac{\alpha_5}{\alpha_3}}\tan\Big[\sqrt{\alpha_5\alpha_3}\xi\Big].
\end{array}
\end{equation}

The above solution is very complicated because the values of $\alpha_3$ and $\alpha_5$ are very complicated. Therefore, we shall study the simple case as follows: $\alpha_0=\frac{d^2}{2c}-\frac{3d}{2}-c$.

Now, using (\ref{0u42-10-2}), (\ref{0u42-8}), (\ref{0u42-6}) and (\ref{u42-1}), and after some calculation, we can obtain the solutions of the metric functions in this case as follows:

$${\textbf{Case~(b-1):}~m>3+\frac{2}{m}}$$

\begin{equation}\label{uu1-a-2-i}
\left\{
  \begin{array}{ll}
A(x,t)=q_1\exp\left[\frac{K\Big[(2m^3-5m^2+m-2)x-(m^2-3m-2)t\Big]}{\sqrt{2}(m-1)^2K_0}\right] \\ ~~~~~~~~~~~~~~ \times \cos^{\gamma_0}[\theta],\\
B(x,t)=q_2 \exp\left[\frac{2\sqrt{2}\gamma_0 Kx}{K_0}\right]\cos^{(m-1)\gamma_0}[\theta],\\
C(x,t)=q_3 \exp\left[\frac{\sqrt{2}\gamma_0 K\Big[(m^2-m-2)x-(m^2-3m-2)t\Big]}{K_0}\right]\\~~~~~~~~~~~~~~ \times  \cos^{(m+1)\gamma_0}[\theta],
  \end{array}
\right.
\end{equation}
where $\gamma_0=\frac{m}{(m-1)^2}$, $K_0^2=m-3-\frac{2}{m}$, $f(x)=c_6\exp\left[\frac{\sqrt{2}K(m^3-3m^2+m-1)x}{K_0(m-1)^2}\right]$ and $\theta=K(x-t)$, the symbols $K$, $m$, $q_1$, $q_2$ and $q_3$ all are being arbitrary constants, however, here $m$ will never be 0 or 1.\\

$${\textbf{Case~(b-2):} m<3+\frac{2}{m}}$$

\begin{equation}\label{uu1-a-2-ii}\left\{
  \begin{array}{ll}
A(x,t)=q_1 \exp\left[\frac{K\Big[(2m^3-5m^2+m-2)x-(m^2-3m-2)t\Big]}{\sqrt{2}K_0(m-1)^2}\right] \\ ~~~~~~~~~~~~~~\times \cosh^{\gamma_0}[\theta],\\
B(x,t)=q_2 \exp\left[\frac{2\sqrt{2}K\gamma_0 x}{K_0}\right]\cosh^{(m-1)\gamma_0}[\theta],\\
C(x,t)=q_3 \exp\left[\frac{\sqrt{2}K\gamma_0 \Big[(m^2-m-2)x-(m^2-3m-2)t\Big]}{K_0}\right]\\ ~~~~~~~~~~~~~~\times \cosh^{(m+1)\gamma_0}[\theta],
  \end{array}
\right.
\end{equation}
where $\gamma_0=\frac{m}{(m-1)^2}$, $K_0^2=\frac{2}{m}+3-m$, $f(x)=c_6 \exp\left[\frac{\sqrt{2}K(m^3-3m^2+m-1)x}{K_0(m-1)^2}\right]$, $\theta=K(x-t)$, the symbols $K$, $m$, $q_1$, $q_2$ and $q_3$ all are being arbitrary constants as above, however, here also $m$ must never be 0 or 1.\\

\section{Validity of the cosmological models: a special case study}
As mentioned in the previous Sec. 3, we are now performing a study regarding different properties of the model in Eq. (50)
under {\it subcase (b-1)}. One can observe that if we take $m =$ 0 or 1, the values of the constants diverse to infinity. For this reason we have purposely skipped the {\bf Case~(b-2)} as this prescription represents a non-realistic model.

For this model, from the equation set (\ref{uu1-a-2-i}), we get the following physical parameters 
\begin{equation}\label{uu1-1a-2-i}
  \begin{array}{ll}
\rho(x,t)=\frac{2\sqrt{2}\gamma_0^2 K^2}{\chi q_1^2 K_0^2}\exp\left[\frac{\sqrt{2}K\Big[(m^2-3m-2)t
-(2m^3-5m^2+m-2)x\Big]}{K_0 (m-1)^2}\right]\\
\times\Big(\sqrt{2}(3+4m-m^3)+K_0(1+2m+3m^2)\tan[\theta]\Big)\cos^{-2\gamma_0}[\theta],
  \end{array}
\end{equation}

\begin{equation}\label{uu1-2a-2-i}
  \begin{array}{ll}
p(x,t)=\frac{2\sqrt{2}\gamma_0^2 K^2}{\chi q_1^2 K_0^2}\exp\left[\frac{\sqrt{2}K\Big[(m^2-3m-2)t
-(2m^3-5m^2+m-2)x\Big]}{K_0(m-1)^2}\right]\\
\times\Big(\sqrt{2}(m^3-2m^2-2m+1)+K_0 (1-2m-m^2)\tan[\theta]\Big)\\
\times\cos^{-2\gamma_0}[\theta],
  \end{array}
\end{equation}

\begin{equation}\label{uu1-3a-2-i}
  \begin{array}{ll}
\frac{F^2_{12}(x,t)}{\bar{\mu}(x,t)}=
\frac{4\sqrt{2}m^2q_2^2 K^2 \exp\left[4\sqrt{2}\gamma_0 Kx/K_0\right]}{\chi (m-1)^4 K_0^2 \cos^{(2-2m)\gamma_0}[\theta]} \times\\
\Big(\sqrt{2}(m^3-2m^2-2m-1)-K_0 (m^2+1)\tan[\theta]\Big),
  \end{array}
\end{equation}
where $\rho$, $p$ and $\bar{\mu}$ are the energy density, pressure  and magnetic permeability respectively with $\gamma_0=\frac{m}{(m-1)^2}$, $\theta=K(x-t)$.

Now, by considering the condition (\ref{eq8}), we have
\begin{equation}\label{uu2-1}
  \begin{array}{ll}
\bar{\mu}(x,t)=\frac{4K^2 \gamma_0^2 q_3^2}{K_0^2 \chi f(t)} \exp\left[\frac{2 \sqrt{2}K \gamma_0 \Big[(m^2-m-2)x
-(m^2-3m-2)t\Big]}{K_0}\right]\\
\times\Big(2m^3-4m^2-4m-2-\sqrt{2}K_0 (m^2-1) \tan[\theta]\Big) \cos^{2 \gamma_0(m+1)}\left[\theta\right],
  \end{array}
\end{equation}

\begin{equation}\label{uu2-2}
  \begin{array}{ll}
F_{12}(x,t)=-\frac{4K^2 \gamma_0^2 q_3 q_2}{K_0^2 \chi \sqrt{f(t)}} \exp\left[\frac{\sqrt{2} K\gamma_0 \Big[(m^2-m) x
-(m^2-3m-2)t\Big]}{K_0}\right]\\
\times\Big(2m^3-4m^2-4m-2-\sqrt{2}K_0 (m^2-1)\tan[\theta]\Big)\cos^{2m \gamma_0}\left[\theta\right],
  \end{array}
\end{equation}
\begin{figure}[t!]\centering
\includegraphics[width=7.0cm]{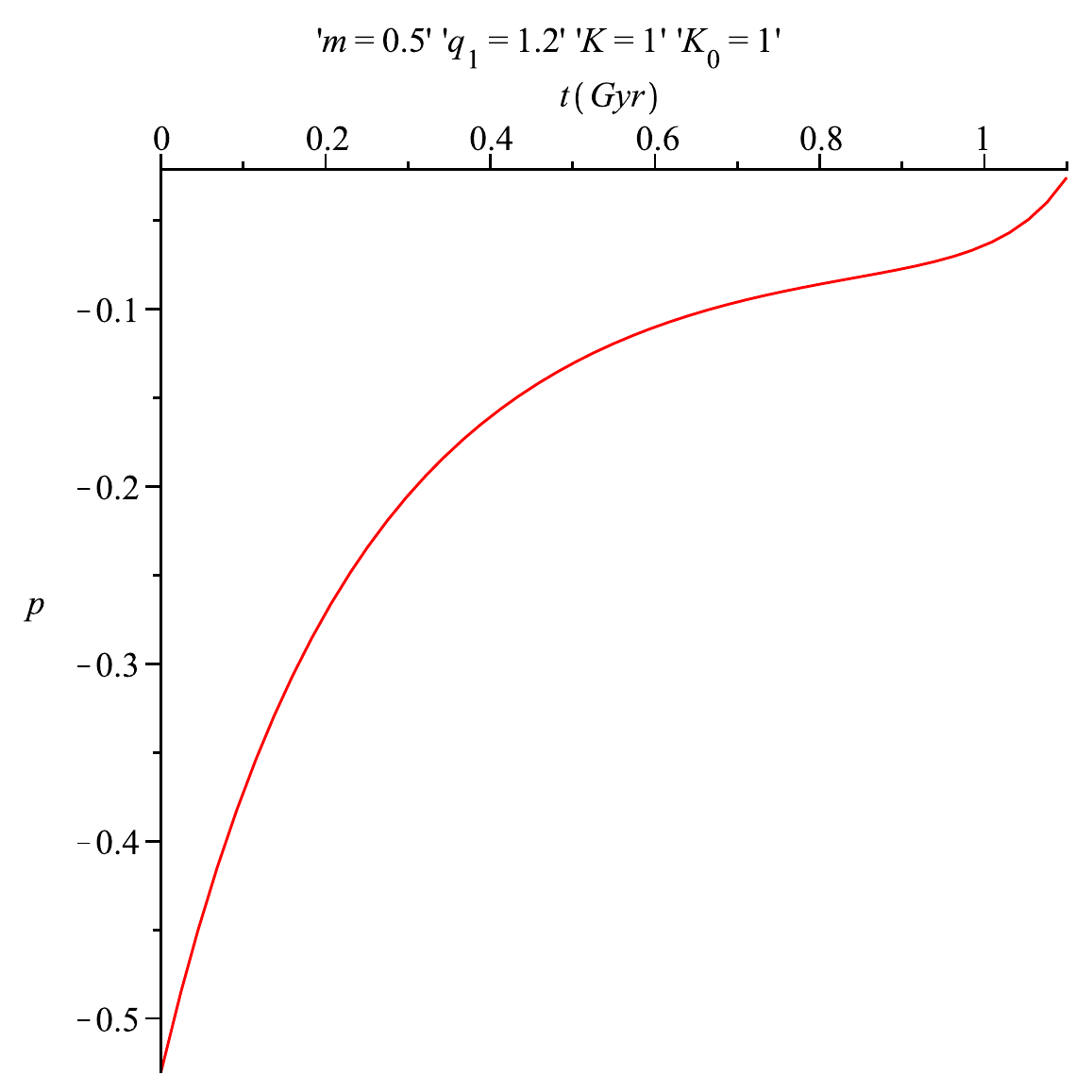}
\includegraphics[width=7.0cm]{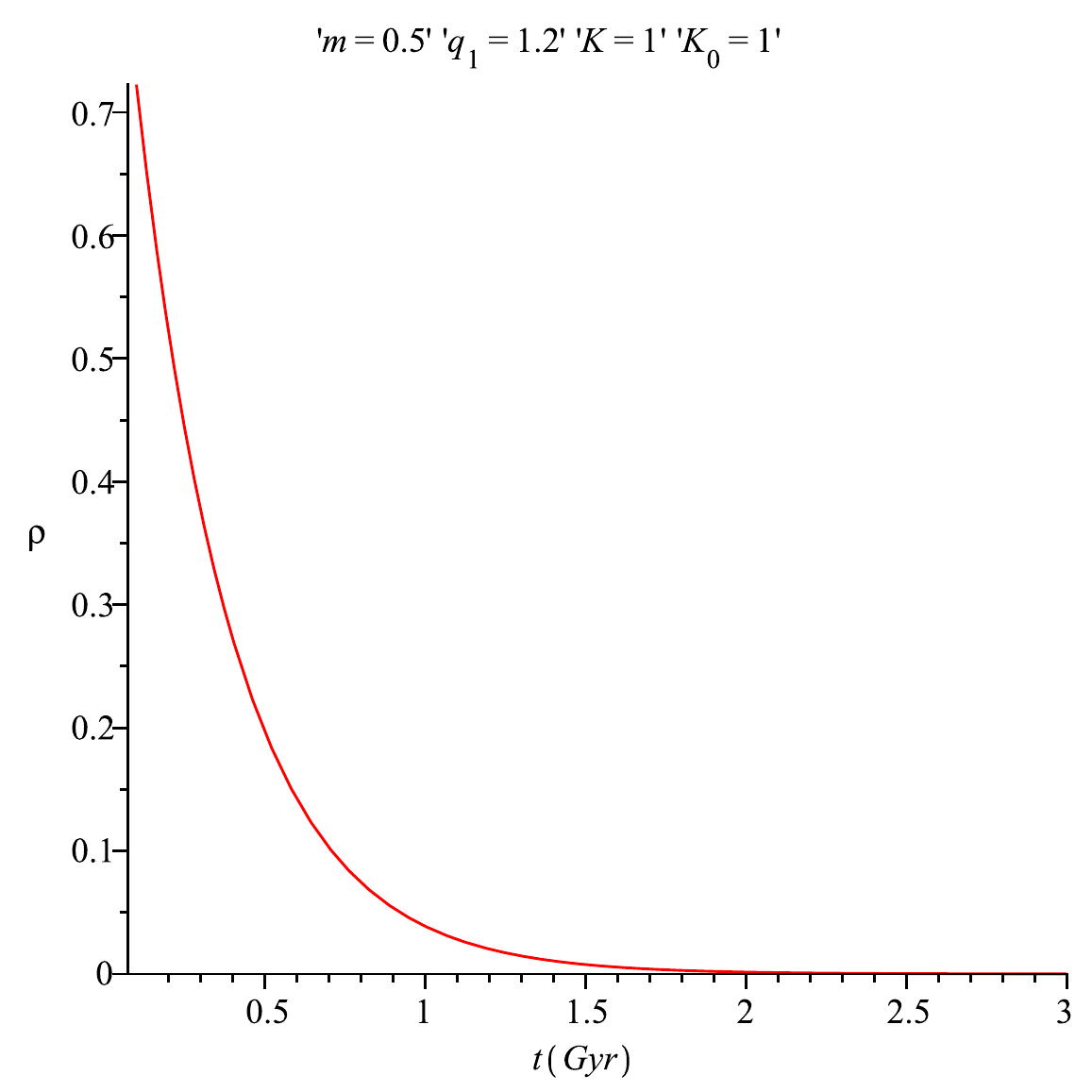}
\caption{Variation of the fluid pressure and matter-energy density with respect to the age of the universe $t$ for the specified values of the constants of the model.}
\label{fig1}
\end{figure}
In Fig. 1 we have drawn the behaviour for $p$ and $\rho$ which show the expected evolutionary features of the universe.\\
\begin{figure}[t!]\centering
\includegraphics[width=7.0cm]{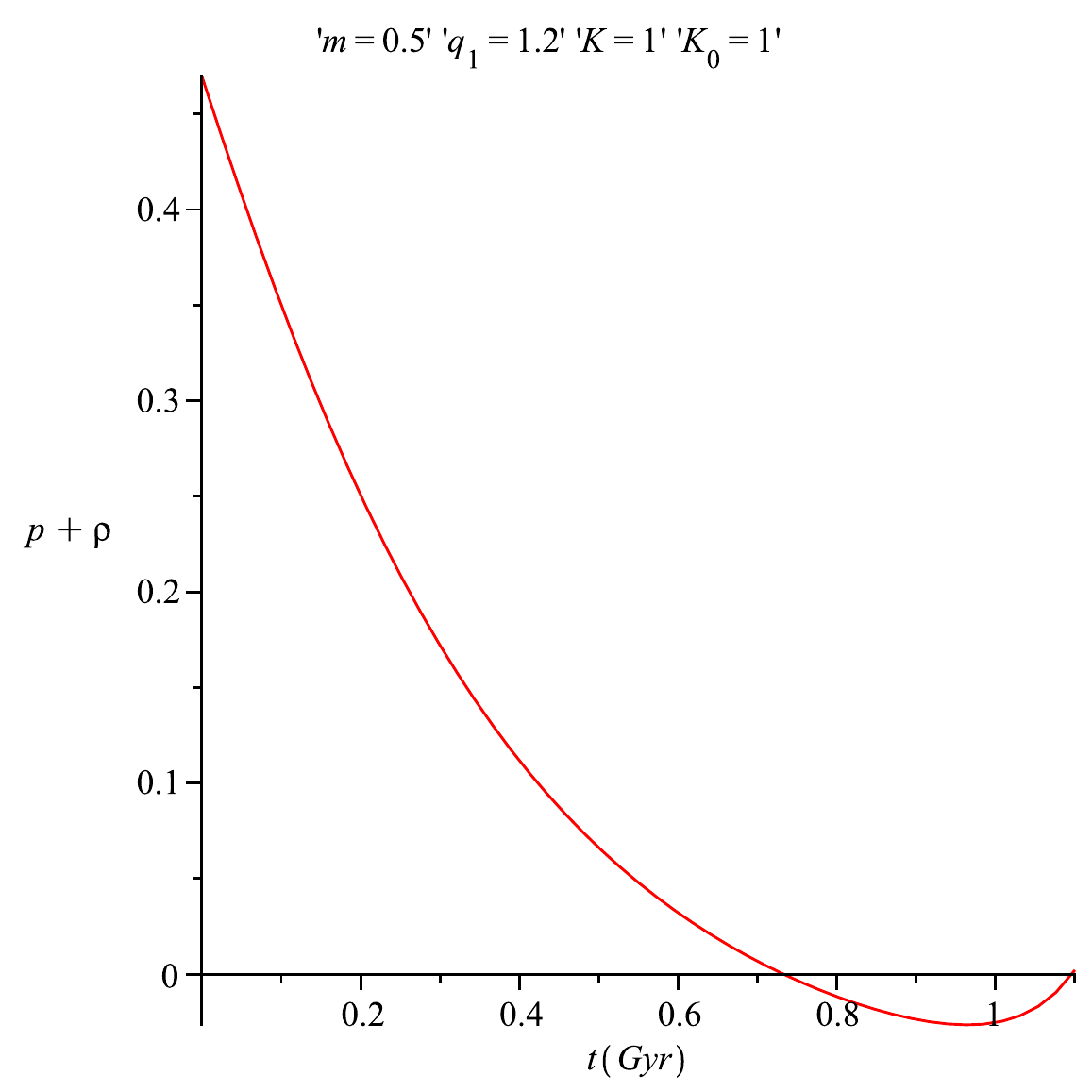}
\caption{Variation of the null energy condition $p+\rho$ with respect to the age of the universe $t$ for the specified values of the constants of the model.}
\label{fig2}
\end{figure}
By using the expressions of density (Eq. \ref{uu1-1a-2-i}) and pressure (Eq. \ref{uu1-2a-2-i}) we also draw plot for $p+\rho$ in Fig. 2. This figure indicates that the null energy condition (i.e. $\rho + p \geq 0$) is obeyed by the system in the early time, however violates at the later stage which supports a deceleration to acceleration feature of the universe.

Now, the expressions for the volume element ($V$), expansion scalar ($\Theta$), shear tensor ($\sigma_i^j$) and shear scalar ($\sigma^2$) are given by
\begin{eqnarray}  \label{uu1-5a-2-i}
V=q_1^2 q_2 q_3~\cos^{(2m+2) \gamma_0}[\theta]~\exp\left[\frac{\sqrt{2}K \Big[(2+5m+2m^2-m^3)t-(2-m+6m^2-3m^3)x\Big]}{(m-1)^2 K_0}\right].
\end{eqnarray}

\begin{equation}\label{uu1-6a-2-i}
  \begin{array}{ll}
\Theta=\frac{(2m+1)\gamma_0 K\Big(\sqrt{2}\tan[\theta]-K_0\Big)}{\sqrt{2} q_1 \cos^{\gamma_0}[\theta]}
\exp\left[\frac{K\Big[(m^2-3m-2)t-(2m^3-5m^2+m-2)x\Big]}{\sqrt{2} K_0 (m-1)^2}\right].
  \end{array}
\end{equation}

\begin{equation}\label{uu1-7a-2-i}
\left\{
  \begin{array}{ll}
\sigma_1^1=\left(\frac{1}{1+2m}-\frac{1}{3}\right)\Theta,\\
\\
\sigma_2^2=\frac{1}{3+6m}\left(\frac{(2+7m+5m^2-2m^3)-\sqrt{2}(m^2-4m)K_0 \tan[\theta]}{
(m^2-3m-2)-\sqrt{2}mK_0 \tan[\theta]}\right)\Theta,\\
\\
\sigma_3^3=-\Big(\sigma_1^1+\sigma_2^2\Big).
  \end{array}
\right.
\end{equation}

\begin{equation}\label{uu1-10a-2-i}
  \begin{array}{ll}
\sigma^2=\frac{m^2-3m-2}{6(1+2m)^2}\left(\frac{\delta_0+\delta_1 \cos[2\theta]-2\sqrt{2}m(m^2+m+1)K_0 \sin[2\theta]}{
\Big[(m^2-3m-2)\cos[\theta]-\sqrt{2}mK_0 \sin[\theta]\Big]^2}\right)\Theta^2,
  \end{array}
\end{equation}
where $\delta_0=4m^4-12m^3-5m^2+9m-2$ and $\delta_1=4m^4-16m^3+3m^2-7m-2$.

On the other hand, non-vanishing acceleration and rotation components are computed as
\begin{equation}\label{uu1-11a-2-i}
\left\{
  \begin{array}{ll}
\dot{u}_1=\frac{K}{\sqrt{2}(m-1)^2 K_0}\Big(\delta_2-\sqrt{2}mK_0 \tan[\theta]\Big),\\

\omega_{41}=-\omega_{14}=\frac{q_1 K\Big(\delta_2-\sqrt{2}mK_0 \tan[\theta]\Big)}{\sqrt{2}(m-1)^2 K_0} 
\times \exp\left[\frac{K\Big[\delta_2 x-(m^2-3m-2)t\Big]}{\sqrt{2}(m-1)^2 K_0}\right]\cos^{\gamma_0}[\theta],
  \end{array}
\right.
\end{equation}
where $\delta_2=2m^3-5m^2+m-2$.

Following the work of Feinstein and lbanez~\cite{Feinstein1993} and Raychaudhuri~\cite{Raychaudhuri1979} the deceleration parameter is given by
\begin{equation}\label{uu1-12a-1}
  \begin{array}{ll}
\mathbf{q}=\frac{\gamma_0 (1+2m)^3 K^4 \Big(\delta_3+\sqrt{2}mK_0 \tan[\theta]\Big)^2}{4q_1^4 (1-m)^5 K_0^2}
\exp\left[\frac{2\sqrt{2}K\Big[\delta_3 t-\delta_2 x\Big]}{(m-1)^2 K_0}\right]\cos^{-2-4\gamma_0}[\theta]\\
\times \Big[m^2-7m+4+(m^2-5m-2)\cos[2\theta]-2\sqrt{2}mK_0 \sin[2\theta]\Big],
  \end{array}
\end{equation}
where $\delta_3=m^2-3m-2$.


\begin{figure}[t!]\centering
\includegraphics[width=7.0cm]{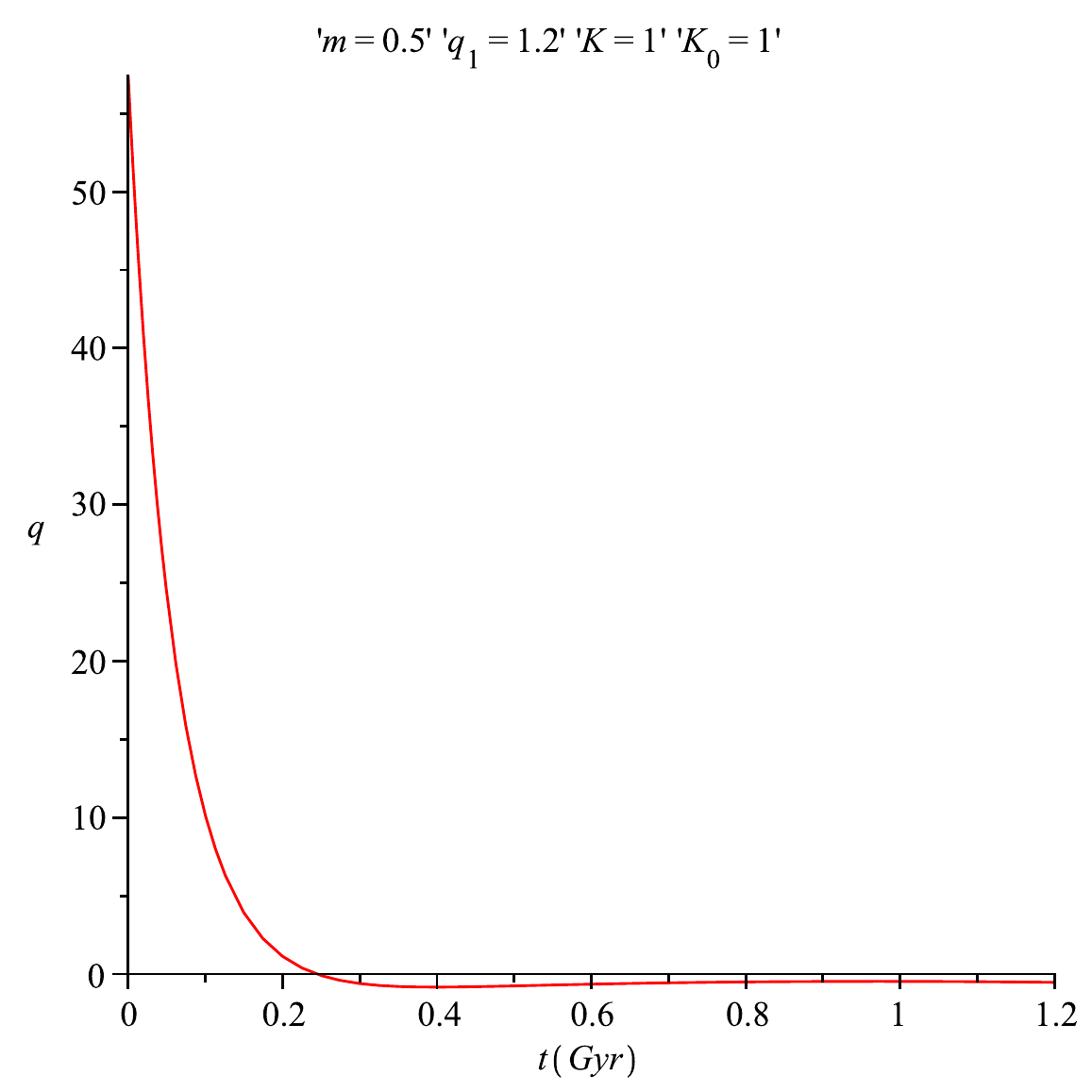}
\caption{Variation of the deceleration parameter $q$ with respect to the age of the universe $t$ for the specified values of the constants of the model.}
\label{fig3}
\end{figure}


The deceleration parameter $q$ is ploted in Fig. 3 which interestingly indicates a change over from the positive $q$ to the negative $q$ with evolution of the universe, i.e. in physical sense from deceleration to accelerating universe. Therefore, the figure actually reveals two particlar features: (i) there is a flip-flop which indicates a slow rolling down of the phase of universe from dceleration to acceleration, and (ii) the phase of acceleration from deceleration has been started from around $t=0.29$~Gyr. In the present epoch of an accelerating universe, $q$ lies near $-0.50 \pm 0.05$ (see the following works~\cite{Tripp1997,Overduin1998,Efstathiou1998,Sahni1999}. From our model, we can recover $q= -0.5$ for $t=0.244$~Gyr when deceleration to acceleration occurs whereas we got $q= -0.5$ at $t=0.29$~Gyr after fine tuning it. However, this data for time seems very low as literature suggests a probable much higher value for $t$ as $\sim 6$~Gyr~\cite{Riess1998,Perlmutter1998,Spergel2003,Kirshner2003,Tegmark2004,Ray2007}.\\
\section{Conclusion}
In the present study, the Lie symmetry analysis has been executed under the Einstein's general relativistic background and hence construction of models for the accelerating universe with perfect fluid and electromagnetic field has been done in plane symmetric spacetime. 

Some interesting and viable physical features of the studies are as follows:

(1) In the present investigation the free gravitational field is assumed to be of the Petrov type-II non-degenerate which provides physically interesting results.

(2) The study deals with the electromagnetic energy of the inhomogeneous universe. The electromagnetic field tensor ($F_{12}$) is found to be positive and increasing function of time.

(3) Among the models presented in Sec. 3 only the case studied in Sec. 4 is found to be interesting with temporial behaviour as far as plots and data are concerned. Other models are with unrealistic physical features having either positive density and volume decreasing or volume increasing but density is negative.

(4) The deceleration parameter $q$ as ploted in Fig. 3 interestingly indicates a change over from positive $q$ to negative $q$ with evolution of the universe i.e. from deceleration to accelerating universe. From our model, we obtain presently accepted numerical value of $q$ as -0.5 for $t=0.24$~Gyr. However, this value of age seems very low with respect to $t \sim 6$~Gyr as available in literature~\cite{Riess1998,Perlmutter1998,Spergel2003,Kirshner2003,Tegmark2004,Ray2007}.

As a final comment we would like to put our overall observations of the present study as follows: qualitatively (see Figs. 1-3) the model under plane-symmetric Einstein-Maxwell spacetime is very promising though quantitative result ($q$ from Fig. 3) seems does not fit for the observed data. This readily indicates that either the analysis under plane symmetric spacetime is not fully compatible with the observable universe or probably we have missed some of the threads in our whole consideration which are responsible to make the analysis partially compatible.
\section*{Acknowledgments}
SR is grateful to the IUCAA, Pune as well as IMSc, Chennai, India for providing Visiting Associateship under which a part of this work was carried out. FR is also grateful to the IUCAA, Pune, India for providing Visiting Associateship scheme and to SERB, DST for financial support. AM is thankful to DST for financial support under INSPIRE scheme.

\end{document}